\documentstyle[11pt,newpasp,twoside,epsfig]{article}
\begin{document}

\title{On the influence of radio galaxies on cooling flows: M87 and Cygnus A}
\author{A. S. Wilson, A. J. Young and D. A. Smith}
\affil{Astronomy Department, University of Maryland, College Park, MD 20742}

\setcounter{page}{111}
\index{Wilson, A. S.}
\index{Young, A. J.}
\index{Smith, D. A.}

\begin{abstract}
Results from Chandra studies of the intracluster medium around M87 and 
Cygnus A are summarized. 
\end{abstract}

\section{Introduction}

In most clusters of galaxies, the X-ray surface brightness rises steeply
towards the center. The radiative cooling time of the gas typically drops
below 10 Gyr at 70 - 150 kpc and 1 Gyr at 5 - 30 kpc from the center (Fabian
2002). X-ray spectra do show declines in the gas temperature
with decreasing radius within these central regions. These observational
results provide the
primary support for the existence of cooling flows. As the gas loses its energy
to radiation, the
weight of the overlying gas produces a subsonic infall. 
In the traditional picture, non-uniformity of the
density is argued to cause denser, cooler matter to form cold clouds or stars
in the flow, and these are supposed to be deposited over an extended region in 
the center of
the cluster (Fabian 1994). Most published models of cooling flows are 
steady-state ones.

The
major technological advances achieved by Chandra and XMM are enabling much more
critical scrutiny of cooling flows, including not only whether the intracluster
medium (ICM) is, in fact, multiphase
but also their relationship to the radio source often
associated with the central galaxy. 
The latter relationship bears on the long controversy over the importance 
of heating processes
on the intracluster medium (contrast Fabian 2002 and Binney 2002).
M87 (Virgo A) and Cygnus A were among
the first identified extragalactic radio sources and are located near the
centers of rich clusters. In this note, we summarise some of our ongoing
work with Chandra in which we attempt to shed light on the interaction
between these radio sources and the ICM.

\section{M87}

\begin{figure}[t]
\centerline{
\psfig{figure=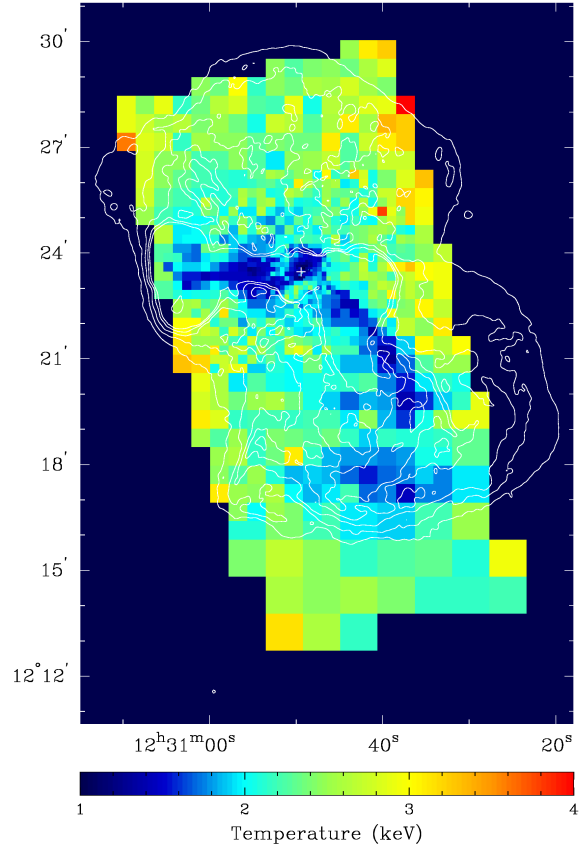,width=8cm,angle=0}
}
\caption{Temperature map of the core of the Virgo cluster (from Young et al.
2002).
This map was constructed from adaptively binned Chandra images.
Overlayed are contours
of 90~cm radio emission (made from VLA archival data).} 
\label{fig1}
\end{figure}

A Chandra study of the X-ray core of the Virgo cluster is given in Young,
Wilson \& Mundell (2002). The main conclusions are as follows.

The inner radio lobes are aligned with depressions in the X-ray surface
brightness and there is no evidence of shock heating in the X-ray emission
immediately surrounding the inner radio lobes, suggesting that the radio plasma
has gently pushed aside the X-ray emitting gas. These cavities cannot have been
inflated much slower than the sound speed, however, or they would have risen
too far from the nucleus due to buoyancy effects. We estimate the jet power to
be L$_{\rm jet} \simeq 3 \times 10^{42}$ erg s$^{-1}$.

On larger scales, the most striking feature is the X-ray arc running from the
east, across the central regions of M87, and off to the southwest (Fig. 1). 
The gas in
the arc has at least two temperatures, with one component at the temperature of
the ambient ICM and a cooler component at
$\simeq 1$ keV. The
gas in the arcs is probably over-pressured with respect to, and somewhat more
metal rich than, the ambient ICM.

Abrupt changes in surface brightness, or ``fronts'', are seen at nuclear
distances slightly larger than the nuclear distances of the inner radio lobes
and intermediate radio ridges. Within the inner front, at nuclear distances
$\le 45^{\prime\prime}$ ($\le 3.5$ kpc) the gas has at least two
temperatures, with the cooler component at $\simeq 1$ keV, similar to the X-ray
arc. This cooler region is concentrated more to the north than the south of the
nucleus and is correlated with the H$\alpha$ + [NII] emission-line
distribution.

We suggest that a model based on the hydrodynamical simulations of
Reynolds, Heinz \& Begelman (2002), scaled
to a lower power radio source such as M87, may explain the observed phenomena.
Intermittent jet activity has two effects on the cluster. Firstly, it inflates
buoyant bubbles of radio plasma that trail cold gas from the central regions in
their wakes as they rise at $\simeq 0.6$ -- 0.7 times the sound speed, thereby
producing the X-ray arcs. The gas dredged up from the nucleus is expected to
have higher metal abundances than the ICM at large nuclear
distances. Secondly, at late times in the evolution of the radio source, the
injection of energy into the cluster core produces a ``pulse'' that expands at
the sound speed into the ICM. 
We suggest that such pulses produce the observed X-ray ``fronts''.
Detailed numerical simulations tailored to the Virgo cluster are required to
explore this hypothesis.

\section{Cygnus A}

Discussions of Chandra results on the radio hot spots, the galactic nucleus
and the large scale ICM are given by Wilson, Young \& Shopbell (2000),
Young et al. (2002) and Smith et al. (2002). Fig. 2 shows the inner regions of 
the X-ray image with temperatures indicated for
various regions of thermally-emitting gas. 
We may bear in mind that an extrapolation inward of the 
temperature of the ICM indicates a temperature at the edge of the cavity of
about 4.7 keV (Smith et al. 2002, improved by better calibration). It is 
striking that the temperatures measured in the bands of emission which project
across the minor axis of the prolate spheroidal cavity are lower, in the
range 3.8 - 4.4 keV (red polygons). 
It is tempting to interpret these structures as ``belts''
of gas extending around the equator of the prolate spheroidal cavity. This
gas is cooling while being accreted by the Cygnus A galaxy, is likely falling
through the cavity and may well continue inwards to form the accretion disk 
around the
nuclear black hole.

In contrast, the temperatures of the limb-brightened edges of the cavity are
5.2 - 6.8 keV (blue polygons), which is hotter than
the inward extrapolation of the cluster gas. If we assume that this high
temperature is the result of a strong shock driven into the surrounding
cluster gas by the expanding cavity, it is possible to estimate the power 
needed to drive the expanding cavity, and hence the minimum power of the jets,
as
L$_{\rm jet}$ $\simeq$ 0.5$\rho_{\rm ICM}$V$^{3}_{\rm S}$A, where
$\rho_{\rm ICM}$ is the pre-shock density, V$_{\rm S}$ the shock velocity 
and A the total surface area of the prolate spheroid. Numerically we have
n$_{\rm ICM}$ $\simeq$ 0.02 cm$^{-3}$ (from the density profile of the cluster
- Smith et al. 2002), V$_{\rm S}$ $\simeq$ 2,000 km s$^{-1}$ (to get a 
postshock temperature of 6 keV), A = 4 $\times$ 10$^{47}$ cm$^{2}$,
giving L$_{\rm jet}$ $\simeq$ 6 $\times$ 10$^{46}$ erg s$^{-1}$. This number
exceeds, by almost a factor of 100, the total radio emission of 
Cygnus A
(L$_{\rm R}$ $\simeq$ 7 $\times$ 10$^{44}$ erg s$^{-1}$) and that of the total
cluster X-ray emission (L$_{\rm X}$(2 - 10 keV) $\simeq$ 1 $\times$ 10$^{45}$ 
erg s$^{-1}$). The cooling time of the shocked gas exceeds the age of the
radio source, so the jet is heating up the inner part
of the cluster. This argument is, to our knowledge, the first to estimate the
jet power from measurements of purely thermal processes, a method which should
be inherently more
reliable than previous estimates involving synchrotron radiation.
Nevertheless, our method revolves around the assumption that the temperature 
measured at the limb-brightened edges represents that behind a strong
shock driven by the expanding cavity. We believe the difference between the 
temperature obtained by inward extrapolation of the
temperature of the large-scale 
ICM -- 4.7 keV -- and those measured around the limb-brightened cavity -- 
5.2 to 6.8 keV -- is real. 
If these numbers are taken literally, the shock
may be weak.
A more complete discussion of the cavity will be
published elsewhere (Wilson \& Smith 2003, in preparation).

\begin{figure}[t]
\centerline{
\psfig{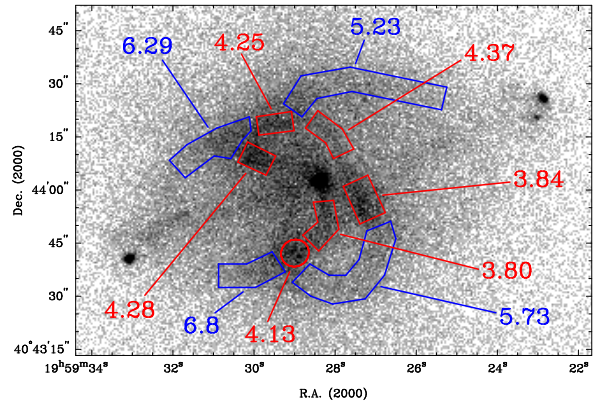}
}
\caption{The Chandra X-ray image of the central regions of Cygnus A. The spectra
of emission from within the indicated polygons have been modelled with a 
MEKAL plasma
and the temperatures (in keV) are given. Red boxes and labelled
temperatures represent the ``bands'', while blue boxes and labelled
temperatures represent the limb-brightened edges of the cavity.
}
\label{fig2}
\end{figure}

\end{document}